# Scalar Quadratic-Gaussian Soft Watermarking Games

M. Kıvanç Mıhçak[2], Emrah Akyol[1], Tamer Başar[1], and Cédric Langbort[1]

[1] Coordinated Science Laboratory, University Of Illinois, Urbana-Champaign, Urbana, IL, 61801, USA
{akyol,basar1,langbort}@illinois.edu
[2] kivancmihcak@gmail.com

**Abstract.** We introduce the zero-sum game problem of *soft watermarking*: The hidden information (watermark) comes from a continuum and has a perceptual value; the receiver generates an estimate of the embedded watermark to minimize the expected estimation error (unlike the conventional watermarking schemes where both the hidden information and the receiver output are from a discrete finite set). Applications include embedding a multimedia content into another. We consider in this paper the scalar Gaussian case and use expected mean-squared distortion. We formulate the resulting problem as a zero-sum game between the encoder & receiver pair and the attacker. We show that for the linear encoder, the optimal attacker is Gaussian-affine, derive the optimal system parameters in that case, and discuss the corresponding system behavior. We also provide numerical results to gain further insight and understanding of the system behavior at optimality.

## 1 Introduction

Watermarking (also termed as information or data hiding throughout the paper) refers to altering an input signal to transmit information in a hidden fashion while preserving the perceptual quality. The watermarked signal is then subject to an attack which "sabotage"s the receiver. In this paper, we focus on "robust" watermarking (unlike steganography or fragile watermarking): The decoder aims to recover the embedded watermark as accurately as possible, even in the presence of (potentially malicious) attacks as long as they preserve the perceptual quality.

Robust watermarking has been an active area of research for nearly two decades, with applications ranging from security-related ones (such as copyright protection, fingerprinting and traitor tracing) to the ones aiming conventional tasks related to multimedia management and processing (such as database annotation, in-band captioning, and transaction tracking). Since the underlying task is to transmit the (hidden) information by means of a watermark, the resulting scheme falls within the category of information transmission problems and can be analyzed using techniques from communications and information theory - see [1] for an overview of data hiding from such a perspective.

Furthermore, the presence of an intelligent attacker enables a game theoretic perspective: Encoder, decoder, and attacker can be viewed as players where the encoder and decoder share a common utility and their gain is exactly the loss of the attacker, thereby resulting in a zero-sum game. In [2, 3], the authors formulated the problem of (robust) information hiding as a game between the encoder & decoder and the attacker; using an information-theoretic approach, they derived expressions for the maximum rate of reliable information transmission (i.e., capacity) for the i.i.d. (independent identically distributed) setup. An analogous approach has been used to extend these results to colored Gaussian signals in [4].

To the best of our knowledge, so far all of the robust watermarking approaches have assumed that the information to be transmitted is an element of a discrete finite (usually binary) set. This is because of the fact that in most intended applications, the watermark is aimed to represent an identity for usage or ownership (or simply permission to use in case of verification problems). Consequently, the receiver is usually designed to decode the embedded watermark (or in case of verification problems detect the presence or absence of a watermark), resulting in a joint source-channel coding problem, where the channel coding counterpart refers to the reliable transmission of the watermark and the source coding counterpart refers to the lossy compression of the unmarked source. In [2–4], following the conventional joint source-channel coding paradigm of information theory, an error event is said to occur if the decoded watermark is not the same as the embedded watermark (hence a hard decision).

In contrast with prior art, in this paper we propose a setup where the information to be transmitted is from a continuum and there is an associated *perceptual value*. As such, the receiver acts as an estimator, whose goal is to produce an estimate of the hidden information from the same continuum (rather than a decoder or detector that reaches a hard decision). Applications include the cases where we hide one multimedia signal inside another (such as embedding one smaller low-resolution image inside another larger high-resolution image, or hiding an audio message inside a video, etc). In such cases, the receiver output is from a continuum as well and there is no hard decision made by it (unlike the prior art in robust watermarking); hence the receiver provides a solution to a *soft decision* problem[3,4]. Accordingly, we use the term "*soft watermarking*" to refer to such data hiding problems. Therefore, unlike the prior art where the fundamental problem is joint source-channel coding, in this case we have a joint *source-source* coding problem where the encoder needs to perform lossy compression on both the unmarked source and the data to be hidden.

---

[3] This distinction is reminiscent of the differentiation between hard and soft decoding methods in classical communication theory.

[4] One alternative approach for this problem may involve using a separated setup, where we first apply lossy compression to the information to be hidden that possesses perceptual value, and subsequently embed the compression output into the unmarked host using a conventional capacity-achieving data hiding code. It is not immediately clear which approach is superior; a comparative assessment constitutes part of our future research.

As a first step toward our long-term goal of studying the soft watermarking problem in its full generality, we consider here a simpler version of the broad soft watermarking problem to gain insight: we confine ourselves to the scalar-Gaussian case where we use expected squared error as the distortion metric. In the future, our aim is to address the asymptotically high dimensional vector case (i.e., information-theoretic setting) with a general class of distributions associated with arbitrary distortion metrics.

In Sec. 2, we introduce the notation and provide the problem formulation. In Sec. 3, we present the main results: In Sec. 3.1, we show that Gaussian-affine attack mapping is optimal for the class of linear encoders; in Sec. 3.2, we derive optimal system parameters for such encoder and attacker classes; in Sec. 3.3, we discuss the system properties at optimality, provide bounds and analyze asymptotic behavior. We present numerical results in Sec. 4 and concluding remarks in Sec. 5.

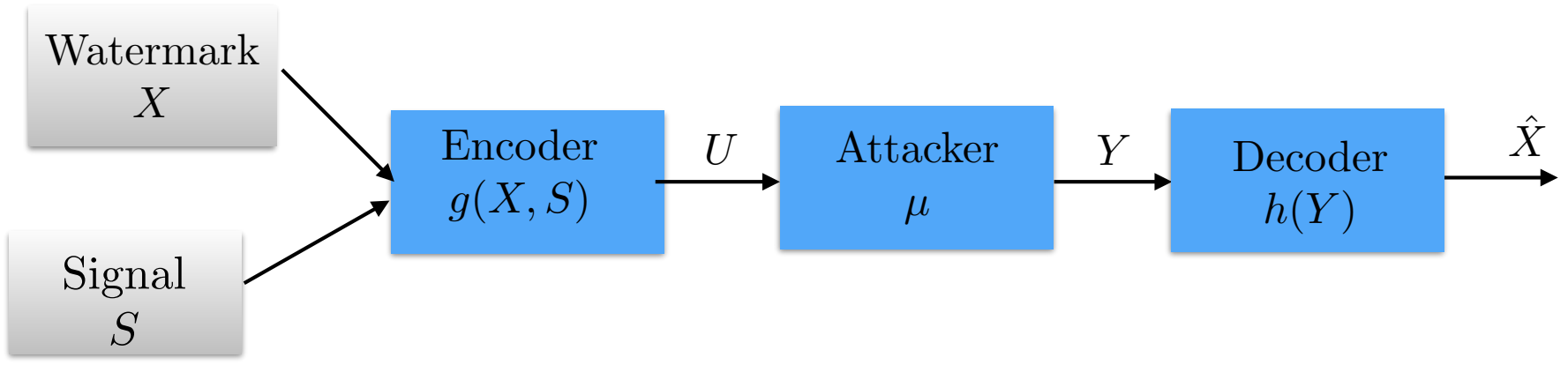

**Fig. 1.** The problem setting.

## 2 Preliminaries

### 2.1 Notation

Let $\mathbb{R}$ and $\mathbb{R}^+$ denote the respective sets of real numbers and positive real numbers. Let $\mathbb{E}(\cdot)$ denote the expectation operator.

The Gaussian density with mean $\mu$ and variance $\sigma^2$ is denoted as $\mathcal{N}\left(\mu, \sigma^2\right)$. All logarithms in the paper are natural logarithms and may in general be complex valued, and the integrals are, in general, Lebesgue integrals.

### 2.2 Problem Definition

A generic model of the problem is presented in Fig. 1. We consider independent scalar Gaussian random variables $X \sim \mathcal{N}\left(0, \sigma_x^2\right)$ and $S \sim \mathcal{N}\left(0, \sigma_s^2\right)$ to denote the watermark (the data to be hidden) and the signal, respectively.

A deterministic mapping of $X$ and $S$ is transmitted over the channel generated by the encoder[5]. Let the transmitter strategy be denoted by $g(\cdot,\cdot)$, which is an element of the space $\Gamma_T$, of real-valued Borel measurable functions satisfying the constraint:

$$\mathbb{E}\left\{(\,g(X,S)-S)^2\right\} \leq P_E. \tag{1}$$

We note that this is a classical constraint that limits the distortion incurred by the watermark embedding process.

The attacker (also termed as the jammer) has access to the output of the transmitter, $U = g(X,S)$, and outputs a random transformation of $U$, denoted by $Y$, i.e., assigns a probability measure, $\mu$, on the output $Y$ that satisfies

$$\int_{-\infty}^{\infty} \mathbb{E}\left\{(Y-U)^2|Y\right\} \,\mathrm{d}\mu(Y) \leq P_A \tag{2}$$

We denote the class of all associated probability measures $\mu$ for the jammer by $\mathcal{M}$. We note that the constraint (2) corresponds to the classical distortion constraint on the attacker used in the watermarking literature (see e.g., [2, Equation 13], [3, Equation 2.2]): It aims to guarantee that the attacker does not distort the watermarked signal beyond a perceptually acceptable level. Thus, in our framework, the attacker has two (possibly conflicting) objectives: i) maximize the distortion between the watermark and its generated estimate by the decoder (cf. (3)), and ii) maintain the usability of the attack output from a perceptual point of view (captured by (2)).

Note that, the constraint (2) differs from the traditional power constraint in classical communication (jamming) games, where the constraint on the attacker arises due to the physical limitations on the communication channel and can be formulated as a power constraint on the attack output $Y$ (i.e., an upper bound on $\mathbb{E}\left(Y^2\right)$) [6]. Since such physical limitations do not exist in our current problem formulation, such a constraint is not immediately applicable to our setup[6]. Also note that, in assessing the perceptual quality of $Y$, as a first step and following [2,3], we take $U$ as the benchmark for the attacker to compare. Alternatively, it is plausible to use $S$ as the benchmark (see e.g., "type-S" constraint in [4, Equation 2.3]), which implies imposing an upper bound on $\mathbb{E}\left(S-Y\right)^2$ and constitutes part of our future research as well.

We consider the power constraints (1,2) in the expectation form, mainly for tractability purposes. Constraints for each realization (in almost sure sense) were also used in the literature [2], but are beyond the scope of our treatment here.

The receiver applies a Borel-measurable transformation $h(Y)$ on its input $Y$, so as to produce an estimate $\hat{X}$ of $X$, by minimizing the squared error distortion

[5] In classical (information-theoretic) watermarking literature, a pseudo-random key sequence is shared between the encoder and the decoder, mainly to render the attacker strategies memoryless. In this paper, we do not consider the key sequence in the problem formulation since our formulation is based on single-letter strategies.

[6] The investigation of a potential relationship between (2) and imposing an upper bound on $\mathbb{E}\left(Y^2\right)$ for the data hiding setup constitutes part of our future research.

measure

$$J\left(g,h,\mu\right)=\int\limits_{-\infty}^{\infty}\mathbb{E}\left\{(h(Y)-X)^2|Y\right\}\,\mathrm{d}\mu(Y) \tag{3}$$

We denote the class of all Borel-measurable mappings $h\left(\cdot\right)$ to be used as the receiver mapping by $\Gamma_R$. Joint statistics of $X$ and $S$, and all objectives are common knowledge.

The common objective of the transmitter and the receiver is to minimize $J$ by properly choosing $g \in \Gamma_T$ and $h \in \Gamma_R$, while the objective of the attacker is to maximize $J$ over the choice of $\mu \in \mathcal{M}$. Since there is a complete conflict of interest, this problem constitutes a zero-sum game.

However, this game does not have a value, since the lower value of the game

$$\bar{J}=\sup_{\mu\in\mathcal{M}}\inf_{\substack{g\in\Gamma_T\\h\in\Gamma_R}}J(g,h,\mu)$$

is not well defined. This is because the attacker cannot guarantee that (2) is satisfied without knowing the encoder strategy $g$ since the value of the left hand side of (2) depends on the joint distribution of $U$ and $Y$ which is impacted by $g$[7]. Hence our main focus is on the following minimax optimization problem which corresponds to the upper value of the game (which in fact safeguards the transmitter-receiver pair against worst attacks and is more relevant for the problem at hand)

$$J^*=\inf_{\substack{g\in\Gamma_T\\h\in\Gamma_R}}\sup_{\mu\in\mathcal{M}}J(g,h,\mu). \tag{4}$$

Note that, the aforementioned formulation implies that the encoder mapping $g\left(\cdot\right)$ is known by the attacker. Note also that for each $g \in \Gamma_T$, we have a zero-sum game between the attacker and the receiver. This subgame has a well-defined value[8], and hence, inf and sup operators can be interchanged, and further they can be replaced by min and max, respectively, i.e., (4) is equivalent to

$$J^*=\inf_{g\in\Gamma_T}\max_{\mu\in\mathcal{M}}\min_{h\in\Gamma_R}J(g,h,\mu), \tag{5}$$

which we consider throughout the rest of the paper. In other words, there is no loss for the encoder-decoder team to determine and announce the decoder mapping before the attacker picks its own mapping, or there is no gain for the decoder to know the attacker mapping *a priori.*

[7] One way to get around this problem is to introduce soft constraints into the objective of the attacker. Then, the problem is no longer a zero-sum game. Another way is to define the attacker constraint for each realization, in almost sure sense, in which case the attacker can satisfy its constraint for any encoding strategy. These are beyond the scope of this paper.

[8] This is a zero-sum game where the objective is linear (hence, concave) in the attacker mapping for a fixed decoder map, and the optimal decoder mapping is unique (conditional mean) for a given attacker mapping. $\mathcal{M}$ is weak*-compact, and the minimizing $h$ can be restricted to a compact subset of $\Gamma_R$ (with (3) bounded away from zero); hence a saddle point exists due to the standard min-max theorem of game theory in infinite-dimensional spaces [7].

# 3 Main Results

Given $Y$, the encoder mapping $g(\cdot)$, and the attacker mapping $\mu(\cdot)$, the decoder's goal is to calculate the estimate $\hat{X}(Y)$ of $X$ so as to minimize the expected squared error distortion measure (cf. (3)) $J = \mathbb{E}\left[X - \hat{X}(Y)\right]^2$. The minimizer here is the well-known MMSE (minimum mean squared error) estimate of $X$ given $Y$: $\hat{X}_{MMSE}(Y) = \mathbb{E}(X|Y)$. Then, the resulting problem is

$$\inf_{g} \max_{\mu} J = \mathbb{E}\left[X - \hat{X}_{MMSE}(Y)\right]^2 \tag{6}$$

subject to the constraints (1,2). In Sec. 3.1, we show that, in the sense of (6), the optimal functional form of $\mu$ is a "jointly-Gaussian-affine mapping" provided that the functional form of the encoder is linear. Using this result, in Sec. 3.2 we solve the problem (6) within the class of linear encoder mappings subject to (1,2)[9] and characterize the parameters of the corresponding system. In Sec. 3.3, we provide a detailed discussion on the optimality results presented in Sec. 3.2.

## 3.1 On Optimal Functional Forms of the Encoder and Attacker

We first focus on a special case as an auxiliary step, where the encoder mapping is the identity operator, $g(X,S) = X$, together with a "generalized" version of the constraint (2), where an upper bound is imposed on $\mathbb{E}(Y - aU)^2$ for an arbitrary $a \in \mathbb{R}$. We present the corresponding optimality result of this special case in Lemma 1. We then use Lemma 1 as an auxiliary step to reach the main result of this section, which states that the Gaussian-affine attack mapping is optimal for the class of linear encoders under the constraint (2); this is reported in Lemma 2.

**Lemma 1.** *Given the encoder strategy of $U = g(X,S) = X$, the solution to*

$$\max_{\mu} \mathbb{E}\left[X - \hat{X}_{MMSE}(Y)\right]^2 \tag{7}$$

*subject to an upper bound on $\mathbb{E}(Y - aU)^2$ for some $a \in \mathbb{R}$ is of the form $Y = \kappa U + Z$ where $Z \sim \mathcal{N}\left(0, \sigma_z^2\right)$ is independent of $U$.*

*Proof.* Define $\mathcal{C}(m_{XY}, m_{YY}) \triangleq \left\{\mu \,\middle|\, \mathbb{E}_\mu(XY) = m_{XY}, \mathbb{E}_\mu\left(Y^2\right) = m_{YY}\right\}$, where $\mathbb{E}_\mu(\cdot)$ denotes expectation with respect to the joint distribution of $X$ and $Y$, induced by the attack mapping $\mu(\cdot)$. Next, define $\hat{X}'(Y) \triangleq \frac{\mathbb{E}(XY)}{\mathbb{E}(Y^2)} Y$. Thus, by

[9] As such, the result provided in Sec. 3.2 forms an upper bound on the solution of (6); see Remark 1 in Sec. 3.3 for a further discussion.

definition, $\hat{X}'(Y) := \hat{X}'_{\mathcal{C}}(Y)$ is the same for all $\mu \in \mathcal{C}(m_{XY}, m_{YY})$ given a pair $(m_{XY}, m_{YY})$. This further implies that, for any given $\mu \in \mathcal{C}(m_{XY}, m_{YY})$,

$$\mathbb{E}\left[X - \hat{X}_{MMSE,\mu}(Y)\right]^2 \leq \mathbb{E}\left[X - \hat{X}'_{\mathcal{C}}(Y)\right]^2 \tag{8}$$

by the definition of the MMSE estimate. The upper bound of (8) is achieved when $X$ and $Y$ are jointly Gaussian, in which case $\hat{X}_{MMSE} = \hat{X}_{LMMSE} = \hat{X}'_{\mathcal{C}}$, which is a well-known result. Thus, we conclude that among all the attack strategies which yield the same $\mathbb{E}(XY)$ and $\mathbb{E}(Y^2)$, the one (if exists) that renders $X$ and $Y$ jointly Gaussian achieves the maximum cost, thereby being the optimal choice for the attacker. A similar reasoning was used in Lemma 1 of [8] for a zero-delay jamming problem.

Next, note that, for any given $\mu \in \mathcal{C}(m_{XY}, m_{YY})$ and $\sigma_x^2 = \mathbb{E}(X^2)$, we have

$$\mathbb{E}_\mu [Y - aU]^2 = \mathbb{E}_\mu [Y - aX]^2 = m_{YY} - 2am_{XY} + \sigma_x^2,$$

implying that all elements of $\mathcal{C}(m_{XY}, m_{YY})$ yield the same $\mathbb{E}(Y - aU)^2$.

Let $\mu^*$ be a solution to (7) subject to an upper bound on $\mathbb{E}(Y - aU)^2$. Assuming existence, let $\mu'$ be an element of $\mathcal{C}\left(\mathbb{E}_{\mu^*}(XY), \mathbb{E}_{\mu^*}(Y^2)\right)$ that renders $X$ and $Y$ jointly Gaussian. Due to the aforementioned arguments, we have $\mathbb{E}\left[X - \hat{X}_{MMSE,\mu^*}(Y)\right]^2 = \mathbb{E}\left[X - \hat{X}_{MMSE,\mu'}(Y)\right]^2 = \mathbb{E}\left[X - \hat{X}'_{\mathcal{C}}(Y)\right]^2$ and that $\mathbb{E}_{\mu^*}[Y - aU]^2 = \mathbb{E}_{\mu'}[Y - aU]^2$. Hence, if $\mu'$ exists, it is optimal.

Existence: Consider the mapping $Y = \kappa U + Z = \kappa X + Z$ for some $\kappa$ and $Z \sim \mathcal{N}(0, \sigma_z^2)$ independent of $X$. Then, straightforward algebra reveals that $\kappa = \mathbb{E}(XY)/\sigma_x^2$ and $\sigma_z^2 = \mathbb{E}(Y^2) - [\mathbb{E}(XY)]^2/\sigma_x^2$. Hence, given $\sigma_x^2$, there is a one-to-one mapping between the pairs of $\left(\mathbb{E}(XY), \mathbb{E}(Y^2)\right)$ and $(\kappa, \sigma_z^2)$. Consequently, for any given $\left(\mathbb{E}(XY), \mathbb{E}(Y^2)\right)$, we can find $(\kappa, \sigma_z^2)$ that guarantees the existence of $\mu'$ that renders $X$ and $Y$ jointly Gaussian. This completes the proof. □

**Lemma 2.** *Given the linear encoder strategy of $U = g(X, S) = \alpha X + \beta S$ for some $\alpha, \beta \in \mathbb{R}$, the solution to*

$$\max_\mu \mathbb{E}\left[X - \hat{X}_{MMSE}(Y)\right]^2, \tag{9}$$

*subject to $\mathbb{E}(Y - U)^2 \leq P_A$, is of the form $Y = \kappa U + Z$, where $Z \sim \mathcal{N}(0, \sigma_z^2)$ is independent of $U$.*

*Proof.* Let $T$ denote the MMSE estimate of $X$ given $U$: $T \triangleq \hat{X}_{MMSE}(U)$. First, note that for any attack mapping $\mu$ and for any function $p(\cdot)$, $(X - T)$ is orthogonal to $p(Y)$:

$$\begin{aligned}\mathbb{E}[(X - T(U))\,p(Y)] &= \mathbb{E}_U\left\{\mathbb{E}\left[Xp(Y)\middle|U\right] - \mathbb{E}\left[T(U)\,p(Y)\middle|U\right]\right\} \\ &= \mathbb{E}_U\left\{\mathbb{E}\left[X\middle|U\right]\mathbb{E}\left[p(Y)\middle|U\right] - T(U)\,\mathbb{E}\left[p(Y)\middle|U\right]\right\} \quad (10) \\ &= 0, \quad (11)\end{aligned}$$

where (10) follows from the fact that $X \leftrightarrow U \leftrightarrow Y$ forms a Markov chain in the specified order, and (11) follows from recalling that $\mathbb{E}\left[X\middle|U\right] = T(U)$ by definition. This implies

$$J = \mathbb{E}\left[X - \hat{X}_{MMSE}(Y)\right]^2 = \mathbb{E}\left[X - T(U) + T(U) - \hat{X}_{MMSE}(Y)\right]^2$$
$$= \mathbb{E}\left[X - T(U)\right]^2 + \mathbb{E}\left[T(U) - \hat{X}_{MMSE}(Y)\right]^2, \tag{12}$$

where (12) follows from the fact that the estimation error $(X - T)$ is orthogonal to any function of $U$ and $Y$ (cf. (11)). Since $\mathbb{E}\left[X - T(U)\right]^2$ is invariant in $\mu$, maximizing $J$ over $\mu$ is equivalent to maximizing $\mathbb{E}\left[T(U) - \hat{X}_{MMSE}(Y)\right]^2$ over $\mu$. Furthermore, since $U$ is linear in $X$ and they are jointly Gaussian, MMSE coincides with LMMSE, implying that $\theta T(U) = U$ for some $\theta \in \mathbb{R}$. Therefore, (9) is equivalent to

$$\max_{\mu} \mathbb{E}\left[T - \hat{X}_{MMSE}(Y)\right]^2, \tag{13}$$

subject to $\mathbb{E}\left(Y - \theta T\right)^2 \leq P_A$. By Lemma 1, we know the solution to (13): At optimality, $Y = \kappa' T + Z$ where $Z \sim \mathcal{N}\left(0, \sigma_z^2\right)$ is independent of $T$. But since $T$ is linear in $U$, this is equivalent to the statement of the lemma. Hence the proof. □

### 3.2 Characterization of Optimal System Parameters

Motivated by Lemma 2, throughout the rest of the paper we confine ourselves to the class of linear mappings for the encoder (14) and jointly-Gaussian-affine mappings for the attacker (15):

$$U = g(X, S) = \alpha X + \beta S, \tag{14}$$
$$Y = \kappa U + Z, \tag{15}$$

where $X \sim \mathcal{N}\left(0, \sigma_x^2\right)$, $S \sim \mathcal{N}\left(0, \sigma_s^2\right)$, $Z \sim \mathcal{N}\left(0, \sigma_z^2\right)$ are all independent of each other. The decoder generates the (L)MMSE estimate of $X$ given $Y$:

$$\hat{X}_{LMMSE}(Y) = \hat{X}_{MMSE}(Y) = \mathbb{E}\left(X \mid Y\right) = \frac{\mathbb{E}(XY)}{\mathbb{E}(Y^2)} Y, \tag{16}$$

with the corresponding mean-squared error cost function

$$J := J(g, h, \mu) = \mathbb{E}\left[X - \hat{X}_{LMMSE}(Y)\right]^2 = \mathbb{E}\left(X^2\right) - \left[\mathbb{E}(XY)\right]^2 / \mathbb{E}\left(Y^2\right). \tag{17}$$

Using (14,15,17) in (1,2,5), the resulting equivalent problem to (5) is given by

$$\min_{\alpha, \beta \in \mathbb{R}} \quad \max_{\kappa \in \mathbb{R}, \sigma_z^2 \in \mathbb{R}^+} \quad J \tag{18}$$
$$\text{s.t.} \qquad \mathbb{E}\left(U - S\right)^2 \leq P_E, \tag{19}$$
$$\mathbb{E}\left(Y - U\right)^2 \leq P_A, \tag{20}$$

where we have replaced "inf" with "min", since $g$ is restricted to linear maps and the cost function is bounded from below by zero (thus restricting $g$ to a compact set without any loss of generality). Note that, the parameters $\alpha, \beta$ (resp. the parameters $\kappa, \sigma_z^2$) constitute the degrees of freedom for the encoder (resp, the attacker) given its linear (resp. affine) nature. Also, (19) (resp. (20)) represents the power constraint for the encoder (resp. the attacker), equivalent to (1) (resp. (2)) to ensure the perceptual fidelity of the marked signal $U$ (resp. the attacked signal $Y$) under the aforementioned parametrization. In Theorem 1, we provide the solution to the minimax problem (18) under the constraints (19,20). Our results are given using the parametrization via $\sigma_u^2 \triangleq \mathbb{E}\left(U^2\right)$. A summary of the results of Theorem 1 is given in Table 1.

**Theorem 1.** *The solution to the minimax soft watermarking problem (18) subject to the constraints (19,20) is as follows:*
(**a**) *For $P_A \le \left(\sigma_s + \sqrt{P_E}\right)^2$, at optimality $\sigma_u^2$ is the unique positive root of the depressed cubic polynomial*

$$f\left(\sigma_u^2\right) \triangleq \sigma_u^6 - \sigma_u^2 \left[\left(\sigma_s^2 - P_E\right)^2 + 2P_A\left(\sigma_s^2 + P_E\right)\right] + 2P_A\left(\sigma_s^2 - P_E\right)^2, \qquad (21)$$

*in the interval of $\left[\max\left(P_A, \left(\sigma_s - \sqrt{P_E}\right)^2\right), \left(\sigma_s + \sqrt{P_E}\right)^2\right]$. The corresponding optimal values of the system parameters are*

$$\beta = \frac{1}{2}\frac{\sigma_u^2 + \sigma_s^2 - P_E}{\sigma_s^2}, \qquad (22)$$

$$\alpha = \sqrt{\frac{\left[\left(\sigma_s + \sqrt{P_E}\right)^2 - \sigma_u^2\right]\left[\sigma_u^2 - \left(\sigma_s - \sqrt{P_E}\right)^2\right]}{4\sigma_s^2\sigma_x^2}}, \qquad (23)$$

$$\kappa = 1 - \frac{P_A}{\sigma_u^2}, \qquad (24)$$

$$\sigma_z^2 = P_A\kappa. \qquad (25)$$

*leading the corresponding optimal value of the cost function as*

$$J = \sigma_x^2 - \frac{\sigma_x^4\alpha^2\left(\sigma_u^2 - P_A\right)}{\sigma_u^4} \qquad (26)$$

(**b**) *If $P_A > \left(\sigma_s + \sqrt{P_E}\right)^2$, then at optimality we have $\kappa = 0$, $\sigma_z^2 \in \left[0, P_A - \sigma_u^2\right]$ is arbitrary, where*

$$\sigma_u^2 = \alpha^2\sigma_x^2 + \beta^2\sigma_s^2 < P_A$$

*for any $\alpha, \beta \in \mathbb{R}$ such that $\alpha^2\sigma_x^2 + (\beta - 1)^2\sigma_s^2 \le P_E$. In that case, the corresponding value of the cost function is given by $J = \sigma_x^2$.*

*Proof.* We first characterize the cost function $J$ (cf. (17)) and the power constraints (19,20) under the given parameterization. Given (14,15) and the inde-pendence of $X$, $S$, $Z$, we have

$$\mathbb{E}(XY) = \kappa\mathbb{E}(XU) = \kappa\alpha\sigma_x^2, \quad (27)$$

$$\mathbb{E}\left(Y^2\right) = \mathbb{E}(\kappa U + Z)^2 = \kappa^2\sigma_u^2 + \sigma_z^2. \quad (28)$$

Using (27,28) in (17), we get

$$J = \sigma_x^2 - \sigma_x^4 \frac{\kappa^2\alpha^2}{\kappa^2\sigma_u^2 + \sigma_z^2}. \quad (29)$$

Furthermore, using (14,15), we can rewrite (19,20) as

$$\mathbb{E}(U - S)^2 = \mathbb{E}[\alpha X + (\beta - 1)S]^2 = \alpha^2\sigma_x^2 + (\beta - 1)^2\sigma_s^2 \leq P_E, \quad (30)$$

$$\mathbb{E}(Y - U)^2 = \mathbb{E}[(\kappa - 1)U + Z]^2 = (\kappa - 1)^2\sigma_u^2 + \sigma_z^2 \leq P_A, \quad (31)$$

where

$$\sigma_u^2 = \mathbb{E}(U)^2 = \mathbb{E}[\alpha X + \beta S]^2 = \alpha^2\sigma_x^2 + \beta^2\sigma_s^2. \quad (32)$$

Step 1 (Inner Optimization): For any given fixed $\alpha, \beta$, we focus on the innermost maximization problem of (18) subject to the constraint (31). Using (29), we observe that an equivalent problem is

$$\min_{\kappa\in\mathbb{R}, \sigma_z^2\in\mathbb{R}^+} J_A, \quad (33)$$

subject to (31), where $J = \sigma_x^2 - \sigma_x^4 J_A$, and

$$J_A \triangleq \frac{\kappa^2\alpha^2}{\kappa^2\sigma_u^2 + \sigma_z^2}. \quad (34)$$

First, consider the special case of $\kappa = 0$: In that case, the attacker erases all the information about the original signal and the watermark and we would have $J_A = 0$. Since $J_A$ is lower-bounded by zero by definition, this would be the optimal policy for the attacker as long as (31) can be satisfied for $\kappa = 0$. In that case, we have $\mathbb{E}(Y - U)^2 \big|_{\kappa=0} = \sigma_u^2 + \sigma_z^2$ per (31). Since the attacker can choose $\sigma_z^2$ arbitrarily small but cannot alter $\sigma_u^2$, we arrive at

$$\text{if } \alpha, \beta \in \mathbb{R} \text{ are such that } P_A \geq \sigma_u^2, \text{ then}$$
$$\kappa_{opt} = 0,\ \sigma_{z,opt}^2 \in \left[0, P_A - \sigma_u^2\right] \text{ is arbitrary, and } J_{A,opt} = 0,\ J_{opt} = \sigma_x^2. \quad (35)$$

Thus, for the rest of the proof, we consider $P_A < \sigma_u^2$ and $\kappa \neq 0$. In that case, we can rewrite $J_A = \frac{\alpha^2}{\sigma_u^2 + \sigma_z^2/\kappa^2}$, and an equivalent problem to (33) is

$$\max_{\kappa\in\mathbb{R}\backslash\{0\}, \sigma_z^2\in\mathbb{R}^+} J_A', \quad (36)$$

subject to (31) where $J_A = \frac{\alpha^2}{\sigma_u^2 + J_A'}$ and $J_A' \triangleq \sigma_z^2/\kappa^2$. Next, note that, for all $\kappa \in \mathbb{R}\backslash\{0\}$, $J_A'$ and left hand side of (31) are both monotonic increasing in $\sigma_z^2$. Thus, the constraint (31) is active at optimality, which yields

$$\sigma_z^2 = P_A - (\kappa - 1)^2\sigma_u^2. \quad (37)$$

Using (37) and the re-parametrization of $t \triangleq 1/\kappa$ in the definition of $J'_A$, we get

$$J'_A = \frac{\sigma_z^2}{\kappa^2} = \frac{P_A - (\kappa-1)^2\sigma_u^2}{\kappa^2} = P_A t^2 - (t-1)^2\sigma_u^2 = \left(P_A - \sigma_u^2\right)t^2 + 2\sigma_u^2 t - \sigma_u^2. \tag{38}$$

Since $P_A < \sigma_u^2$, $J'_A$ is concave in $t$. Hence, maximization of (38) subject to (37) admits a unique solution, given by

$$t_{opt} = -\frac{\sigma_u^2}{P_A - \sigma_u^2} \quad \Leftrightarrow \quad \kappa_{opt} = \frac{\sigma_u^2 - P_A}{\sigma_u^2}, \tag{39}$$

which is equivalent to (24). Using this result in (37) yields

$$\sigma_{z,opt}^2 = P_A - (\kappa_{opt}-1)^2\sigma_u^2 = P_A - \frac{P_A^2}{\sigma_u^2}, \tag{40}$$

which is equivalent to (25). Note that, $\sigma_u^2 > P_A$ implies the positivity of $\kappa_{opt}$ and $\sigma_{z,opt}^2$ per (39) and (40), respectively. Using (39,40) in the cost function definitions, we get

$$J'_{A,opt} \triangleq J'_A\big|_{\kappa=\kappa_{opt},\sigma_z^2=\sigma_{z,opt}^2} = \frac{\sigma_{z,opt}^2}{\kappa_{opt}^2} = \frac{P_A}{\kappa_{opt}} = \frac{P_A\sigma_u^2}{\sigma_u^2 - P_A}, \tag{41}$$

$$J_E \triangleq J_{A,opt} = J_A\big|_{\kappa=\kappa_{opt},\sigma_z^2=\sigma_{z,opt}^2} = \frac{\alpha^2}{\sigma_u^2 + J'_{A,opt}} = \frac{\alpha^2\left(\sigma_u^2 - P_A\right)}{\sigma_u^4}, \tag{42}$$

$$J = \sigma_x^2 - \sigma_x^4 J_E. \tag{43}$$

Note that (26) directly follows from using (42) in (43).
Step 2 (Outer Optimization): Next, given the solution to the inner optimization problem of (18) in (42), we proceed with solving the corresponding outer optimization problem, given by

$$\max_{\alpha,\beta\in\mathbb{R}} J_E, \tag{44}$$

subject to the constraint (30) and $\sigma_u^2 > P_A$.

First, we show that, without loss of generality (w.l.o.g.) we can assume $\alpha, \beta \geq 0$. Define the left hand side of (30) as a bivariate function of $(\alpha, \beta)$:

$$q(\alpha,\beta) \triangleq \alpha^2\sigma_x^2 + (\beta-1)^2\sigma_s^2. \tag{45}$$

Note that both $J_E$ and $q$ are even functions of $\alpha$. So, w.l.o.g. we can assume $\alpha \geq 0$. Furthermore, for any $\beta < 0$, we have $(\beta-1)^2 = (|\beta|-1)^2 + 2|\beta|$. Hence, for any $\alpha \in \mathbb{R}$ and $\beta < 0$, we have $q(\alpha,\beta) = q(\alpha,|\beta|) + 2|\beta|\sigma_s^2$. This implies that for any $\beta < 0$, $[q(\alpha,\beta) \leq P_E] \Rightarrow [q(\alpha,|\beta|) \leq P_E]$. Combining this observation with the fact that $J_E$ is an even function of $\beta$, we reach the conclusion that w.l.o.g. we can assume $\beta \geq 0$.

Next, we show that for $\sigma_u^2 > P_A$, the constraint (30) is active at optimality. In order to do that, we examine the behavior of both $J_E$ and $q$ with respect to $\alpha^2$ (noting that there is a one-to-one mapping between $\alpha$ and $\alpha^2$ since we assume $\alpha \geq 0$ w.l.o.g.). We have

$$\frac{\partial q}{\partial \alpha^2} = \sigma_x^2 > 0, \tag{46}$$

$$\begin{aligned}
\frac{\partial J_E}{\partial \alpha^2} &= \frac{\sigma_u^2 - P_A}{\sigma_u^4} + \alpha^2 \frac{\partial \sigma_u^2}{\partial \alpha^2} \frac{\sigma_u^4 - 2\sigma_u^2\left(\sigma_u^2 - P_A\right)}{\sigma_u^8}, \\
&= \frac{\sigma_u^2 - P_A}{\sigma_u^4} + \alpha^2 \sigma_x^2 \frac{-\sigma_u^2 + 2P_A}{\sigma_u^6}, \\
&= \frac{1}{\sigma_u^6}\left[\left(\sigma_u^2 - P_A\right)\sigma_u^2 - \alpha^2\sigma_x^2\left(\sigma_u^2 - P_A\right) + \alpha^2\sigma_x^2 P_A\right] \\
&= \frac{1}{\sigma_u^6}\left[\left(\sigma_u^2 - P_A\right)\beta^2\sigma_s^2 + \alpha^2\sigma_x^2 P_A\right] > 0,
\end{aligned} \tag{47}$$

where (47) follows from using (32) and recalling that $\sigma_u^2 > P_A$ by assumption. Now, the monotonicity results (46,47) jointly imply that, at optimality the encoder will choose $\alpha$ as large as possible for any fixed $\beta$ provided that the power constraint (30) is satisfied. Therefore, the constraint (30) is active at optimality.

Using the fact that the constraint is active at optimality, we have

$$\begin{aligned}
P_E = \mathbb{E}\left(U - S\right)^2 = q\left(\alpha, \beta\right) &= \alpha^2\sigma_x^2 + \beta^2\sigma_s^2 + \left(1 - 2\beta\right)\sigma_s^2, \\
&= \sigma_u^2 + \left(1 - 2\beta\right)\sigma_s^2,
\end{aligned} \tag{48}$$

where (48) follows from (32). Using (48), we get (22).

Our next goal is to rewrite $J_E$ (cf. (42)) as a function of $\sigma_u^2$, and accordingly formulate the problem (44) as a univariate maximization problem in terms of $\sigma_u^2$. Using (22) in (32), we obtain

$$\sigma_u^2 = \alpha^2\sigma_x^2 + \frac{1}{4}\left(\frac{\sigma_u^2 + \sigma_s^2 - P_E}{\sigma_s^2}\right)^2 \sigma_s^2 = \alpha^2\sigma_x^2 + \frac{\left(\sigma_u^2 + \sigma_s^2 - P_E\right)^2}{4\sigma_s^2}. \tag{49}$$

Using (49), we get

$$\begin{aligned}
\alpha^2 &= \frac{4\sigma_s^2\sigma_u^2 - \left(\sigma_u^2 + \sigma_s^2 - P_E\right)^2}{4\sigma_s^2\sigma_x^2} = -\frac{\left(\sigma_u^2 + \sigma_s^2 - P_E - 2\sigma_s\sigma_u\right)\left(\sigma_u^2 + \sigma_s^2 - P_E + 2\sigma_s\sigma_u\right)}{4\sigma_s^2\sigma_x^2}, \\
&= -\frac{\left[\left(\sigma_u - \sigma_s\right)^2 - P_E\right]\left[\left(\sigma_u + \sigma_s\right)^2 - P_E\right]}{4\sigma_s^2\sigma_x^2}, \qquad (50) \\
&= -\frac{1}{4\sigma_s^2\sigma_x^2}\left[\left(\sigma_u - \sigma_s - \sqrt{P_E}\right)\cdot\left(\sigma_u - \sigma_s + \sqrt{P_E}\right)\cdot\right. \\
&\qquad \left.\left(\sigma_u + \sigma_s - \sqrt{P_E}\right)\cdot\left(\sigma_u + \sigma_s + \sqrt{P_E}\right)\right] \\
&= -\frac{1}{4\sigma_s^2\sigma_x^2}\left[\sigma_u^2 - \left(\sigma_s + \sqrt{P_E}\right)^2\right]\left[\sigma_u^2 - \left(\sigma_s - \sqrt{P_E}\right)^2\right], \qquad (51) \\
&= \frac{-\left[\sigma_u^2 - \left(\sigma_s^2 + P_E\right)\right]^2 + 4P_E\sigma_s^2}{4\sigma_s^2\sigma_x^2}. \qquad (52)
\end{aligned}$$

Using (51), we get (23). Also, the analysis of the constraint $\alpha^2 \geq 0$ yields

$$\left[\alpha^2 \geq 0\right] \Leftrightarrow \left[\left(\sigma_s - \sqrt{P_E}\right)^2 \leq \sigma_u^2 \leq \left(\sigma_s + \sqrt{P_E}\right)^2\right], \tag{53}$$

which directly follows from (51). Moreover, (53) and the constraint $\left(\sigma_u^2 > P_A\right)$ jointly imply that, if $P_A > \left(\sigma_s + \sqrt{P_E}\right)^2$, the feasible set for the problem (44) is empty. In that case, the encoder cannot generate a powerful enough marked signal $U$ such that $\sigma_u^2 > P_A$. Then, at optimality the attacker chooses $\kappa = 0$ and erases $U$ (i.e., (35) is valid). As a result, if $P_A > \left(\sigma_s + \sqrt{P_E}\right)^2$ (which is equivalent to $\sqrt{P_A} - \sqrt{P_E} > \sigma_s$), Theorem 1(b) holds.

On the other hand, if

$$\left[P_A < \left(\sigma_s + \sqrt{P_E}\right)^2\right] \quad \Leftrightarrow \quad \left[\sqrt{P_A} - \sqrt{P_E} < \sigma_s\right], \tag{54}$$

the problem (44) reduces to

$$\max_{\sigma_u^2} J_E\left(\sigma_u^2\right) \tag{55}$$

$$\text{s.t.} \max\left[P_A, \left(\sigma_s - \sqrt{P_E}\right)^2\right] \leq \sigma_u^2 \leq \left(\sigma_s + \sqrt{P_E}\right)^2, \tag{56}$$

where

$$J_E\left(\sigma_u^2\right) = -\frac{\left\{\left[\sigma_u^2 - \left(\sigma_s^2 + P_E\right)\right]^2 - 4P_E\sigma_s^2\right\}\left(\sigma_u^2 - P_A\right)}{4\sigma_s^2\sigma_x^2\sigma_u^4}, \tag{57}$$

which follows from using (52) in (42).

Next we quantify asymptotic behavior of $J_E$ which will be useful in characterizing properties of its extrema. We proceed by first defining the numerator of $J_E$ as $N\left(\sigma_u^2\right)$:

$$N\left(\sigma_u^2\right) \triangleq -\left\{\left[\sigma_u^2 - \left(\sigma_s^2 + P_E\right)\right]^2 - 4P_E\sigma_s^2\right\}\left(\sigma_u^2 - P_A\right),$$

which, after some straightforward algebraic manipulations, leads to

$$\begin{aligned}\frac{\partial N}{\partial \sigma_u^2} &= -\left[\sigma_u^2 - \left(\sigma_s^2 + P_E\right)\right]\left[3\sigma_u^2 - \left(\sigma_s^2 + P_E + 2P_A\right)\right] + 4\sigma_s^2 P_E, \\ \frac{\partial^2 N}{\partial\left(\sigma_u^2\right)^2} &= -6\sigma_u^2 + 4\left(\sigma_s^2 + P_E\right) + 2P_A. \end{aligned} \tag{58}$$

Thus, we have

$$\lim_{\sigma_u^2 \to 0} J_E\left(\sigma_u^2\right) = \lim_{\sigma_u^2 \to 0} \frac{\left(\sigma_s^2 - P_E\right)^2 P_A}{4\sigma_s^2\sigma_x^2\sigma_u^4} \quad \to \quad \infty, \tag{59}$$

$$\lim_{\sigma_u^2 \to \infty} J_E\left(\sigma_u^2\right) = \lim_{\sigma_u^2 \to \infty} \frac{\partial^2 N/\partial\left(\sigma_u^2\right)^2}{8\sigma_s^2\sigma_x^2} \quad \to \quad -\infty, \tag{60}$$

$$\lim_{\sigma_u^2 \to -\infty} J_E\left(\sigma_u^2\right) = \lim_{\sigma_u^2 \to -\infty} \frac{\partial^2 N/\partial\left(\sigma_u^2\right)^2}{8\sigma_s^2\sigma_x^2} \quad \to \quad \infty, \tag{61}$$

where (59) follows from (57), and (60,61) follow from using (58).

Next note that, $J_E\left(\sigma_u^2\right)$ has 3 roots: $P_A$, $\left(\sigma_s - \sqrt{P_E}\right)^2$, $\left(\sigma_s + \sqrt{P_E}\right)^2$. The first one is obvious with a direct inspection of (57); the second and third roots directly follow from noting the equality of (51) and (52), and using that in (57).

Assuming that the feasible set for the problem (44) is non-empty, i.e., (54) holds, (54,59,60,61) jointly imply

$$\begin{gathered}\left\{0 < \sigma_u^2 < \min\left[P_A, \left(\sigma_s - \sqrt{P_E}\right)^2\right]\right\} \Rightarrow J_E\left(\sigma_u^2\right) > 0, \\ \left\{\min\left[P_A, \left(\sigma_s - \sqrt{P_E}\right)^2\right] < \sigma_u^2 < \max\left[P_A, \left(\sigma_s - \sqrt{P_E}\right)^2\right]\right\} \Rightarrow J_E\left(\sigma_u^2\right) < 0, \\ \left\{\max\left[P_A, \left(\sigma_s - \sqrt{P_E}\right)^2\right] < \sigma_u^2 < \left(\sigma_s + \sqrt{P_E}\right)^2\right\} \Rightarrow J_E\left(\sigma_u^2\right) > 0, \\ \left\{\left(\sigma_s + \sqrt{P_E}\right)^2 < \sigma_u^2\right\} \Rightarrow J_E\left(\sigma_u^2\right) < 0.\end{gathered} \tag{62}$$

Thus, there are a total of 3 extrema of $J_E(\cdot)$. The one that is of interest to us, i.e., the one which satisfies (56), is a maximizer by (62). Furthermore, there is a unique such stationary point within the feasible region of (56). In order to calculate this maximizer, we take the derivative of (57) with respect to $\sigma_u^2$. After some straightforward algebra we get $\left(-4\sigma_s^2\sigma_x^2\sigma_u^4\right)\frac{dJ_E}{d\sigma_u^2} = f\left(\sigma_u^2\right)$, where $f(\cdot)$ is a depressed cubic polynomial and is given by (21). The solution of (55) is then given by the unique positive root of (21) which falls in the region specified by (56). Recall that this is the solution if (54) are satisfied. This completes the proof of part (a) of Theorem 1. □

### 3.3 Discussion on the System Behavior at Optimality

**Remark 1 (On Optimality of Theorem 1)** Per Lemma 2, the results reported in Theorem 1 describe the optimal system characterization when the encoder is confined to the class of *linear* mappings (cf. (14)). Hence, in the sense of (17), our results form an *upper bound* on the optimal system performance within an arbitrary class of encoder mappings. Investigation of the tightness of this bound constitutes part of our future research. Throughout the rest of the paper, when we refer to optimality, we mean optimality in the sense of Theorem 1.

**Remark 2 (Trivial and Non-trivial Policies)**
**(a)** We say that "the (optimal) system is trivial" if, at optimality, the attacker erases all the information on the mark-embedded signal $U$ and only retains information on the additive noise $Z$. This coincides with having $Y = Z$ (i.e., $\kappa = 0$). Conversely, we have a "non-trivial system" if the attacked signal $Y$ contains information on the marked signal $U$ (i.e., $\kappa > 0$) at optimality.
**(b)** The case of *non-trivial system* happens if we have $\sigma_u^2 > P_A$ at optimality (due to the arguments leading to (35)). This is possible if and only if $P_A \leq \left(\sigma_s + \sqrt{P_E}\right)^2$ (or equivalently $\sqrt{P_A} - \sqrt{P_E} \leq \sigma_s$) (cf. (54)). In this case, given $\sigma_s^2$ and $P_E$, the encoder is able to design $\alpha$ and $\beta$ such that the power of the marked signal $U$ is larger than the power constraint $P_A$ and is able to transmit information about $X$ through the channel. This case is covered in part (a) of Theorem 1. Conversely, if $P_A > \left(\sigma_s + \sqrt{P_E}\right)^2$ (or equivalently $\sqrt{P_A} - \sqrt{P_E} > \sigma_s$), we have the case of *trivial system*, and it is impossible for the encoder to design $U$ to exceed $P_A$. Then, the optimal attacker can affored to erase $U$, thus essentially sending noise to the decoder. This case is covered in part (b) of Theorem 1.

| Condition | Operation Regime and Cost | Encoder Parameters | Attacker Parameters |
|---|---|---|---|
| $\left[P_A \leq \left(\sigma_s + \sqrt{P_E}\right)^2\right]$ $\Longleftrightarrow$ $\left[\sqrt{P_A} - \sqrt{P_E} \leq \sigma_s\right]$ | Non-Trivial Theorem 1(a) $J = \sigma_x^2 - \frac{\sigma_x^4 \alpha^2 \left(\sigma_u^2 - P_A\right)}{\sigma_u^4}$ | $\sigma_u^2$ unique root of $f\left(\sigma_u^2\right)$ (21) s.t. $\max\left(P_A, \left(\sigma_s - \sqrt{P_E}\right)^2\right) \leq \sigma_u^2 \leq \left(\sigma_s + \sqrt{P_E}\right)^2$, $\alpha$ given by (23), $\beta = \frac{1}{2}\left(1 + \frac{\sigma_u^2 - P_E}{\sigma_s^2}\right)$. | $\kappa = 1 - P_A/\sigma_u^2$, $\sigma_z^2 = P_A \kappa$ . |
| $\left[P_A > \left(\sigma_s + \sqrt{P_E}\right)^2\right]$ $\Longleftrightarrow$ $\left[\sqrt{P_A} - \sqrt{P_E} > \sigma_s\right]$ | Trivial Theorem 1(b) $J = \sigma_x^2$ | Any $\alpha, \beta \in \mathbb{R}$ s.t. $\alpha^2\sigma_x^2 + (\beta - 1)^2 \sigma_s^2 \leq P_E$, $\sigma_u^2 = \alpha^2\sigma_x^2 + \beta^2\sigma_s^2 < P_A$. | $\kappa = 0$, $\sigma_z^2 \in \left[P_A - \sigma_u^2\right]$ . |

**Table 1.** Summary of Theorem 1: Characterization of the scalar-Gaussian soft watermarking system at optimality. The first (leftmost) column indicates the condition that leads to the operation regime specified in the second column (see Remark 2 of Sec. 3.3 for the description and a discussion on "trivial" and "non-trivial" policies); the third and fourth columns show the corresponding values of the encoder and attacker parameters at optimality, respectively. Note that, in case of Theorem 1(b), while there are infinitely many choices of encoder parameters, the choice of $\alpha = 0$, $\beta = 1 + \sqrt{P_E/\sigma_s^2}$ is a sensible one and maintains continuity between regions (see Remark 3 of Sec. 3.3 for details).

**Corollary 1.** ***(Power Constraints)*** *In the non-trivial regime (Theorem 1(a)), the encoder power constraint (19) and the attacker power constraint (20) are both active.*

**Corollary 2.** ***(Cost Ordering)*** *At optimality, the cost of the non-trivial regime is upper-bounded by the cost of the trivial regime,* $\sigma_x^2$.

**Corollary 3.** ***(Existence and Uniqueness)*** *In the non-trivial regime, the optimal system parameters specified in Theorem 1(a) are guaranteed to exist and they are essentially unique.*[10]

[10] In the proof of Theorem 1, all expressions that involve $\alpha$ are, in fact, functions of $\alpha^2$; therefore if $\alpha^*$ is optimal, so is $-\alpha^*$. To account for such multiple trivial solutions, we use the term "essential uniqueness".

Corollaries 1, 2, 3 directly follow from the proof Theorem 1. Specifically, Corollary 1 is a consequence of arguments following (36,47), Corollary 2 follows from using $P_A \leq \sigma_u^2$ in (26), and Corollary 3 is because of (62) and the arguments following it.

**Remark 3 (On Optimal Encoding Parameters in case of the Trivial System)** In case of a trivial system, by Theorem 1(b), there are infinitely many combinations of system parameters that achieve optimality. Among those, one choice for the encoder that is intuitively meaningful is to choose $\alpha = 0$ and $\beta = 1 + \sqrt{P_E/\sigma_s^2}$. This choice corresponds to having $U = \beta S$ such that $\mathbb{E}\left(U - S\right)^2 = \left(\beta - 1\right)^2 \sigma_s^2 = P_E$, i.e., the encoder chooses not to send any information on the watermark $X$ and chooses the scaling factor $\beta$ such that the power constraint (19) is satisfied with equality. Note that, such a choice ensures continuity between the values of the system parameters at optimality across non-trivial and trivial regions (cf. Remark 4).

**Remark 4 (On Optimal Operation Regimes)**
**(a)** Variation with respect to system inputs: For $P_E \geq P_A$, the system always operates at the non-trivial mode at optimality since $\sqrt{P_A} - \sqrt{P_E} \leq 0 \leq \sigma_s$. For fixed $\sigma_s^2$ and $P_E$, as $P_A$ increases, the strength of the attack channel increases and the attacker is able to act with a better "budget". As $P_A$ increases, when we reach $P_A = \left(\sigma_s + \sqrt{P_E}\right)^2 + \epsilon$ for an arbitrarily small $\epsilon > 0$, the feasible region for the encoder becomes the empty set (cf. (56)) and a transition from the non-trivial mode to the trivial mode occurs. Conversely, for fixed $P_A$ and for fixed $P_E$ (resp. $\sigma_s^2$), as $\sigma_s^2$ (resp. $P_E$) increases, we observe a transition from the non-trivial region to the trivial region.
**(b)** Continuity between the regions: Suppose we have a system that is initially in the non-trivial region with $P_A < \left(\sigma_s + \sqrt{P_E}\right)^2$. Then, we have

$$\lim_{P_A \uparrow \left(\sigma_s + \sqrt{P_E}\right)^2} \sigma_u^2 = P_A = \left(\sigma_s + \sqrt{P_E}\right)^2, \tag{63}$$

$$\lim_{P_A \uparrow \left(\sigma_s + \sqrt{P_E}\right)^2} \beta = \frac{1}{2} \frac{\sigma_s^2 - P_E + \left(\sigma_s + \sqrt{P_E}\right)^2}{\sigma_s^2} = 1 + \sqrt{\frac{P_E}{\sigma_s^2}}, \tag{64}$$

$$\lim_{P_A \uparrow \left(\sigma_s + \sqrt{P_E}\right)^2} \alpha = 0, \tag{65}$$

$$\lim_{P_A \uparrow \left(\sigma_s + \sqrt{P_E}\right)^2} \kappa = \lim_{P_A \uparrow \left(\sigma_s + \sqrt{P_E}\right)^2} \sigma_z^2 = 0, \tag{66}$$

$$\lim_{P_A \uparrow \left(\sigma_s + \sqrt{P_E}\right)^2} J = \sigma_x^2, \tag{67}$$

where (63) follows from noting that the feasible region (56) converges to the singleton $P_A = \left(\sigma_s + \sqrt{P_E}\right)^2$, (64,65) follow from using $\sigma_u^2 = \left(\sigma_s + \sqrt{P_E}\right)^2$ in (22,23), respectively, and (66,67) follow from using $\sigma_u^2 = P_A$ in (24,25,26), respectively. Note that, the attack parameters (66) and the optimal cost value

(67) readily satisfy continuity with their unique counterparts of Theorem 1(b). Furthermore, it can be shown that the encoder parameters (64,65) achieve optimality in the trivial regime (cf. Remark 3).

**Remark 5 (Performance Bounds)**
We focus here on deriving bounds on the cost for the more interesting case of Theorem 1(a) when $\left(\sqrt{P_A} - \sqrt{P_E}\right) < \sigma_s$. Note that (26), and problem construction clearly imply the bounds of $0 \leq J \leq \sigma_x^2$. The upper bound is tight and can be attained for $P_A = \left(\sigma_s + \sqrt{P_E}\right)^2$. In order to obtain a potentially tighter lower bound, we initially proceed with deriving an upper bound on $\alpha$. Consider the polynomial

$$g\left(\sigma_u^2\right) \triangleq \left[\left(\sigma_s + \sqrt{P_E}\right)^2 - \sigma_u^2\right]\left[\sigma_u^2 - \left(\sigma_s - \sqrt{P_E}\right)^2\right].$$

Then, it is straightforward to show that $g(\cdot)$ is concave and that

$$\frac{\partial g}{\sigma_u^2} = -2\sigma_u^2 + \left(\sigma_s - \sqrt{P_E}\right)^2 + \left(\sigma_s + \sqrt{P_E}\right)^2 = -2\sigma_u^2 + 2\left(\sigma_s^2 + P_E\right).$$

Hence $g(\cdot)$ is maximized for $\sigma_u^2 = \sigma_s^2 + P_E$. Using this result in (23) yields the following upper bound on $\alpha$:

$$\alpha \leq \sqrt{\frac{g\left(\sigma_u^2\right)\Big|_{\sigma_u^2=\sigma_s^2+P_E}}{4\sigma_s^2\sigma_x^2}} = \sqrt{\frac{P_E}{\sigma_x^2}}. \tag{68}$$

Furthermore, using (68) on (26), we get

$$J \geq \sigma_x^2\left[1 - P_E\frac{\sigma_u^2 - P_A}{\sigma_u^4}\right]. \tag{69}$$

Define $h\left(\sigma_u^2\right) \triangleq \frac{\sigma_u^2 - P_A}{\sigma_u^4}$. It is straightforward to show that for $\sigma_u^2 \geq P_A$, $h(\cdot)$ is concave and the maximizer is $2P_A$, which yields $h\left(\sigma_u^2\right) \leq \frac{1}{4P_A}$. Using this result in (69) and combining it with the previously mentioned bounds, we get

$$\max\left(0, \sigma_x^2\left(1 - \frac{P_E}{4P_A}\right)\right) \leq J \leq \sigma_x^2. \tag{70}$$

We thus have a non-trivial lower bound for $J$ if $P_A > P_E/4$.

**Remark 6 (Role of the Watermark Power $\sigma_x^2$)**
A quick inspection of Theorem 1 reveals that the optimal operational mode of the resulting system depends on $\sigma_s^2$, $P_A$, and $P_E$, and is independent of the watermark power $\sigma_x^2$. Intuitively, this is because of two reasons: First, because of the nature of the underlying problem, we do not impose any distortion constraint between $X$ and any other variable in the system. Next, the scaling parameter $\alpha$ can be used to adjust the contribution of the watermark (cf. (14)) to make it arbitrarily large or arbitrarily small. Indeed, (23) implies the following: For fixed $\sigma_s^2$, $P_A$, $P_E$, if a pair of $\left(\alpha_1, \sigma_{x,1}^2\right)$ is optimal in the sense of Theorem 1, then so is another pair $\left(\alpha_2, \sigma_{x,2}^2\right)$ if and only if $\alpha_1 \sigma_{x,1} = \alpha_2 \sigma_{x,2}$. On the other hand, $\sigma_x^2$ directly affects the value of the resulting cost $J$.

**Remark 7 (Asymptotics - Large Signal Case)**
It is possible to obtain some asymptotics when $\sigma_s^2 \gg P_E, P_A$. In that case, the system will operate at the non-trivial mode at optimality (governed by Theorem 1(a)). Then, (21) can be written as

$$f\left(\sigma_u^2\right) \sim \sigma_u^6 - \sigma_u^2\left(\sigma_s^4 + 2P_A\sigma_s^2\right) + 2P_A\sigma_s^2 = \left(\sigma_u^2 - \sigma_s^2\right)\left(\sigma_u^4 + \sigma_u^2\sigma_s^2 - 2P_A\right). \quad (71)$$

Thus, at optimality, using $\sigma_s^2 \gg P_E, P_A$, we have

$$\sigma_u^2 \sim \sigma_s^2, \quad (72)$$

$$\beta \sim \left[\frac{1}{2}\frac{\sigma_u^2 + \sigma_s^2}{\sigma_s^2}\right] \sim 1, \quad (73)$$

$$\alpha \sim \sqrt{\frac{\left[\left(\sigma_s + \sqrt{P_E}\right)^2 - \sigma_s^2\right]\left[\sigma_s^2 - \left(\sigma_s - \sqrt{P_E}\right)^2\right]}{4\sigma_s^2\sigma_x^2}} = \sqrt{\frac{P_E}{\sigma_x^2}}, \quad (74)$$

$$J \sim \left[\sigma_x^2 - \sigma_x^4\frac{\alpha^2}{\sigma_s^2}\right] \sim \left[\sigma_x^2\left(1 - \frac{P_E}{\sigma_s^2}\right)\right] \quad (75)$$

where (72) follows from the fact that $\sigma_s^2$ is the unique root of (71) in the region of interest, (73) and (74) follow from using (72) in (22) and (23), respectively, (75) follows from using (72,74) in (26).

## 4 Numerical Results

In this section, we numerically illustrate the behavior of the optimal scalar-Gaussian soft watermarking system as a function of the power of the unmarked signal, $\sigma_s^2$. The results are presented in Fig. 2 and Fig. 3 for a fixed watermark power $\sigma_x^2 = 10$ owing to the discussion in Remark 6. Because of the linear relationship between $\kappa$ and $\sigma_z^2$, we do not show $\sigma_z^2$ vs $\sigma_s^2$ plots since $\kappa$ vs $\sigma_s^2$ plots are already present.

In Fig. 2, solid, dashed, dash-dotted and dotted lines correspond to the cases of $(P_A = 1, P_E = 16)$, $(P_A = 1, P_E = 4)$, $(P_A = 4, P_E = 1)$, and $(P_A = 16, P_E = 1)$, respectively. Here,

- (a) and (b) show $J$ vs $\sigma_s^2$ for the whole range and for $\sigma_s^2$ small, respectively;
- (c) and (d) show $\sigma_u^2$ vs $\sigma_s^2$ for the whole range and for $\sigma_s^2$ small, respectively;
- (e), (f), and (g) show $\alpha$, $\beta$ and $\kappa$ as functions of $\sigma_s^2$, respectively.

By Theorem 1, the system always operates in the non-trivial mode for $P_A \leq P_E$ (solid and dashed lines), and it operates in the trivial mode for $\sigma_s^2 \leq$ $\left(\sqrt{P_A} - \sqrt{P_E}\right)^2$ for $P_A > P_E$, i.e., trivial mode for $\sigma_s^2 \leq 1$ for $(P_A = 4, P_E = 1)$ (dash-dotted line) and $\sigma_s^2 \leq 9$ for $(P_A = 16, P_E = 1)$ (dotted line), which is clearly observable in panel (b). Note that, we observe continuity in the behavior of all system elements during the transition between trivial and non-trivial regions.

In Fig. 3, solid and dashed lines represent true values and large-signal approximations for $P_A = P_E = 1$. As expected, the large-signal approximation becomes more accurate as $\sigma_s^2$ increases.

## 5 Conclusions

We have introduced the zero-sum game problem of *soft watermarking* where the hidden information has a perceptual value and comes from a continuum, unlike the prior literature on robust data hiding where the focus has been on cases where the hidden information is an element of a discrete finite set. Accordingly, the receiver produces a soft estimate of the embedded information in the proposed setup. As a first step toward this new class of problems, we focus in this paper on the scalar-Gaussian case with the expected squared estimation error as the cost function and analyze the resulting zero-sum game between the encoder & decoder pair and the attacker. Expected distortion constraints are imposed both on the encoder and the attacker to ensure the perceptual quality. Restricting the encoder mapping to be linear in the watermark and the unmarked host, we show that the optimal attack mapping is Gaussian-affine. We derive closed-form expressions for the system parameters in the sense of minimax optimality. We further discuss properties of the resulting system in various aspects, including bounds and asymptotic behavior, and provide numerical results.

Our future work includes an information-theoretic analysis of the problem considered here, and extensions to settings with privacy and security constraints (see e.g., [5] for analysis of a similar communication problem with privacy constraints).

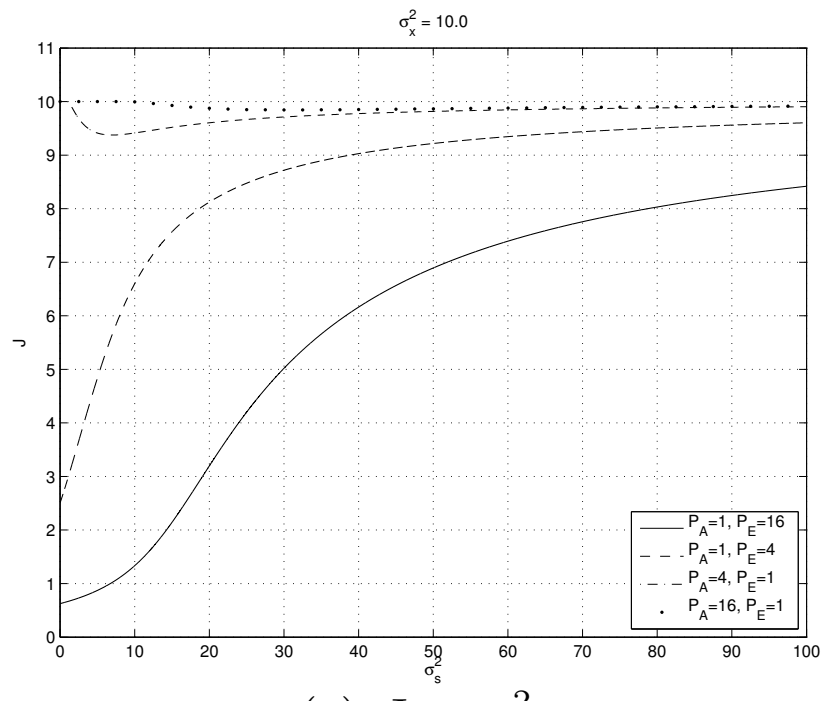

(a) $J$ vs $\sigma_s^2$

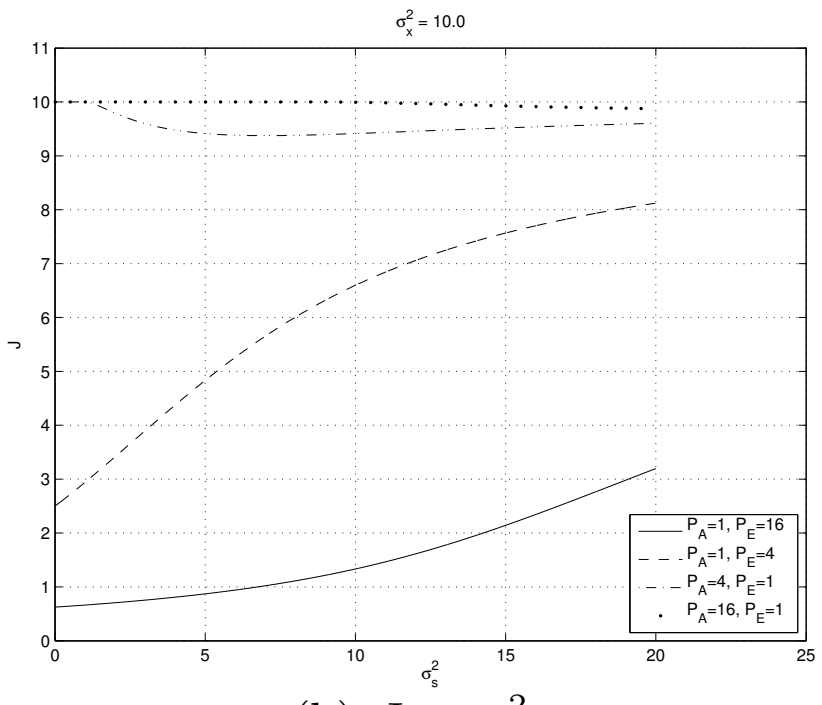

(b) $J$ vs $\sigma_s^2$

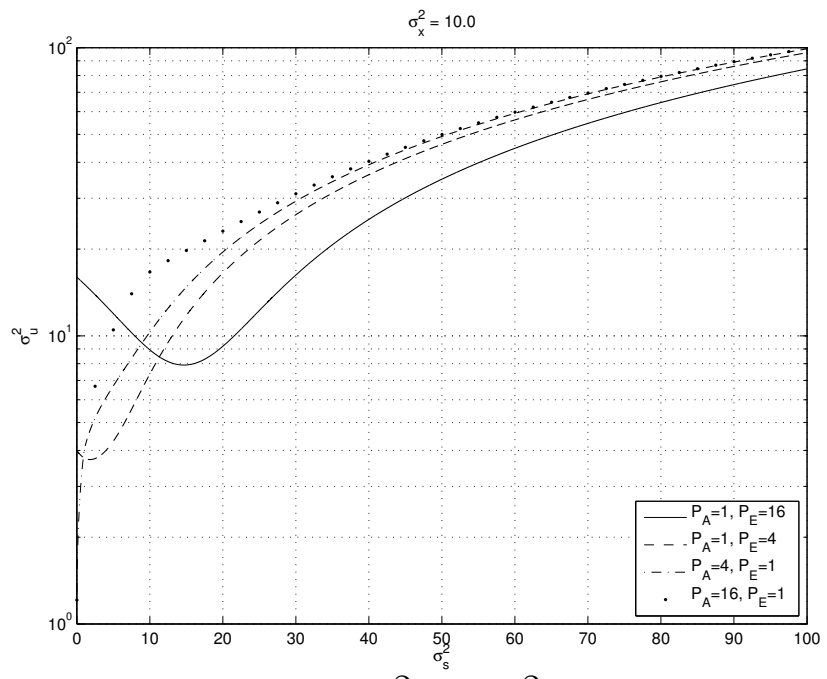

(c) $\sigma_u^2$ vs $\sigma_s^2$

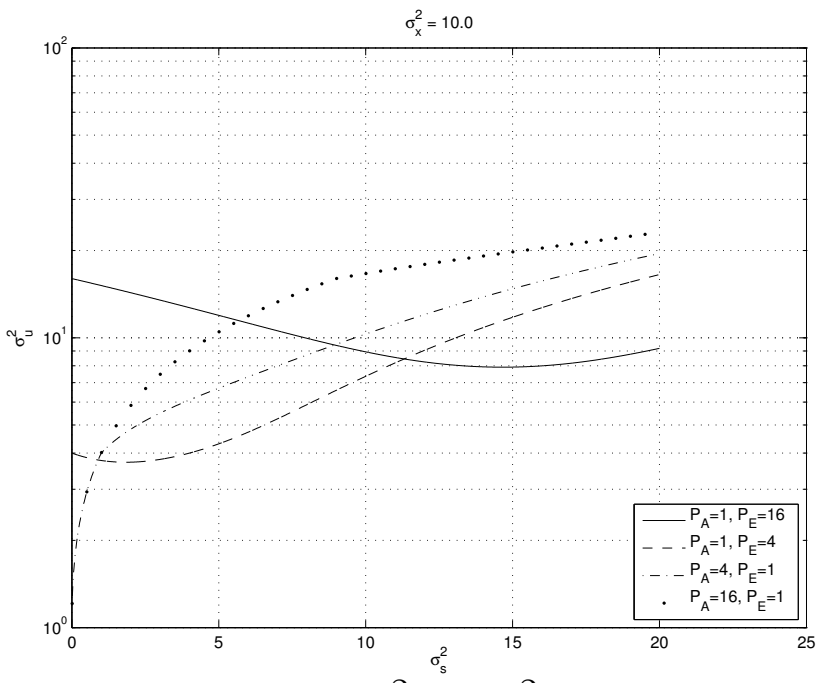

(d) $\sigma_u^2$ vs $\sigma_s^2$

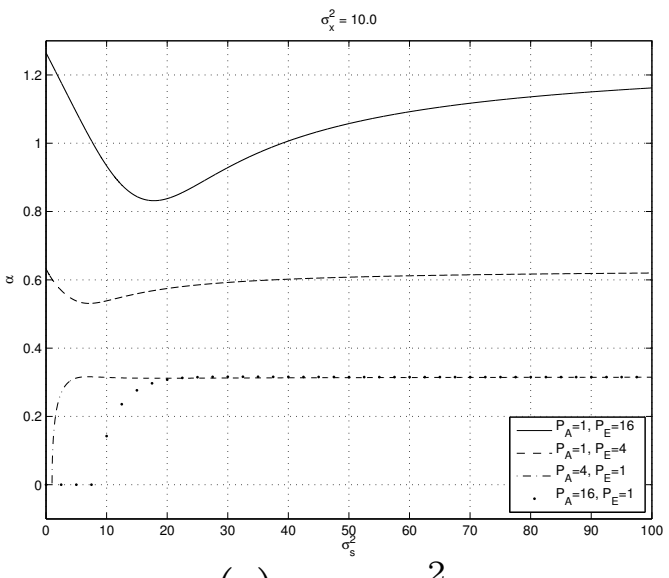

(e) $\alpha$ vs $\sigma_s^2$

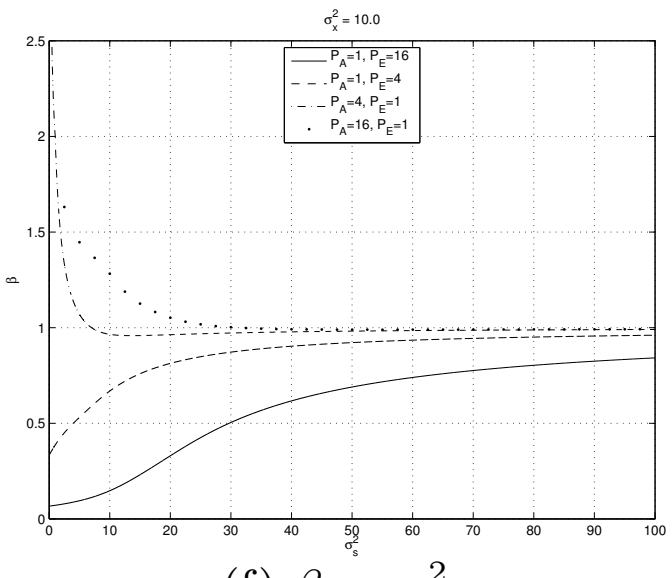

(f) $\beta$ vs $\sigma_s^2$

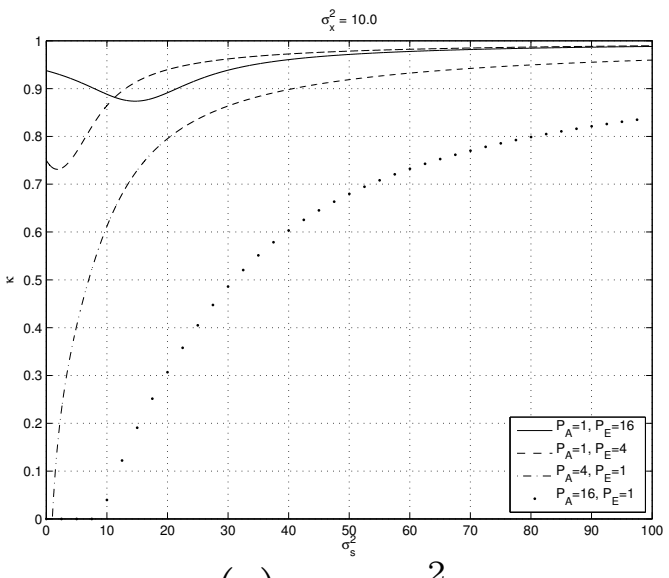

(g) $\kappa$ vs $\sigma_s^2$

**Fig. 2.** System behavior at optimality as a function of $\sigma_s^2$ for the case of $\sigma_x^2 = 10$; solid, dashed, dash-dotted, and dotted lines represent the cases of $(P_A = 1, P_E = 16)$, $(P_A = 1, P_E = 4)$, $(P_A = 4, P_E = 1)$, $(P_A = 16, P_E = 1)$, respectively. By Remark 3 of Sec. 3.3, we use $\alpha = 0$ and $\beta = 1 + \sqrt{P_E/\sigma_s^2}$ in case of a trivial system.

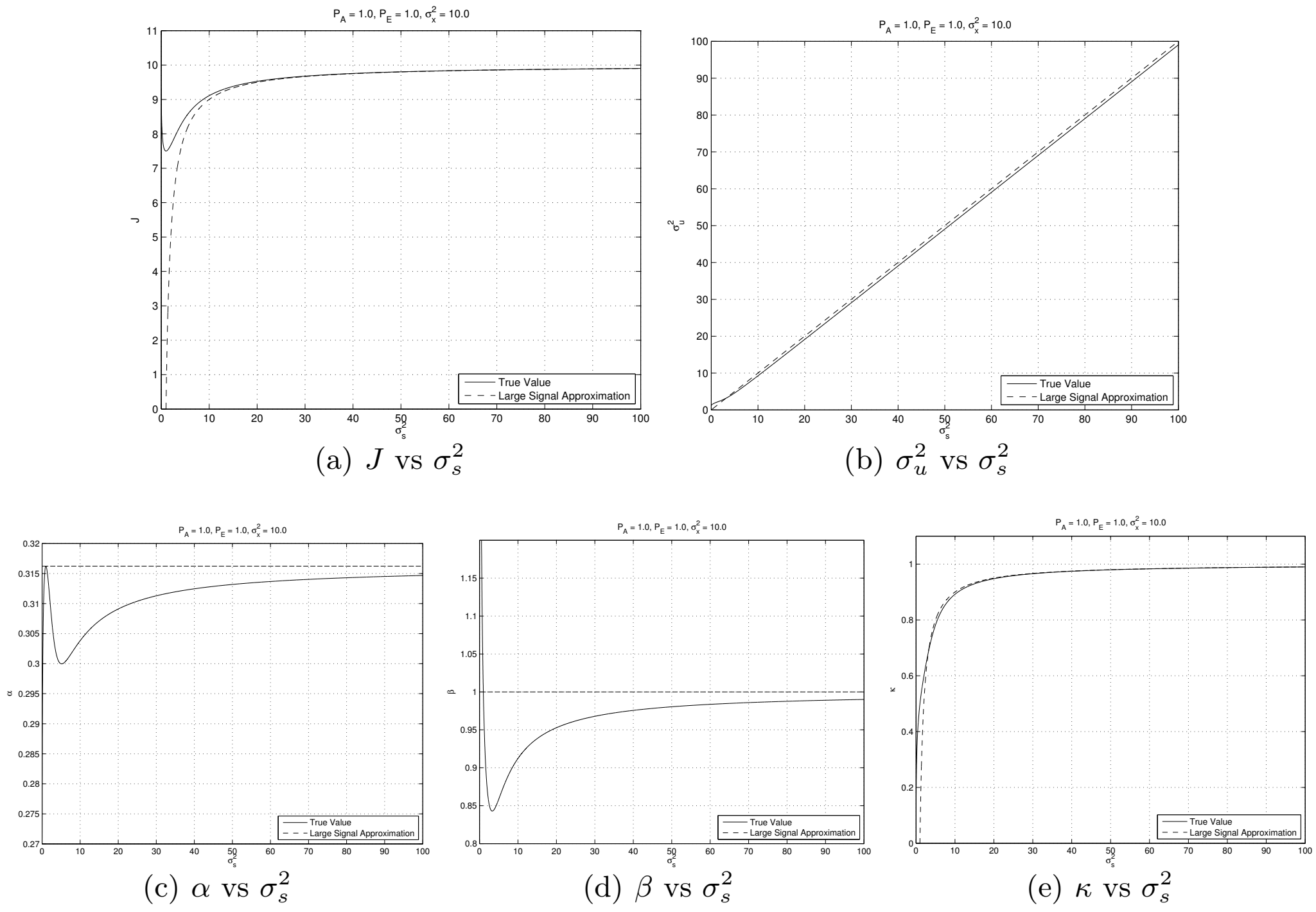

(a) $J$ vs $\sigma_s^2$ (b) $\sigma_u^2$ vs $\sigma_s^2$

(c) $\alpha$ vs $\sigma_s^2$ (d) $\beta$ vs $\sigma_s^2$ (e) $\kappa$ vs $\sigma_s^2$

**Fig. 3.** System behavior at optimality as a function of $\sigma_s^2$ for the case of $\sigma_x^2 = 10$, $P_A = P_E = 1$; solid and dashed lines represent true values and large signal approximation(cf. Remark 7 of Sec. 3.3), respectively.